\documentclass[12pt]{article}

\usepackage{epsfig, amssymb, amsmath}

\DeclareGraphicsExtensions{.jpg, .eps}
\newcommand{\be}{\begin{equation}}
\newcommand{\ee}{\end{equation}}
\newcommand{\bea}{\begin{eqnarray}}
\newcommand{\eea}{\end{eqnarray}}
\newcommand{\ba}{\begin{array}}
\newcommand{\ea}{\end{array}}

\def\bbox{{\,\lower0.9pt\vbox{\hrule \hbox{\vrule height 0.2 cm
\hskip 0.2 cm \vrule height 0.2 cm}\hrule}\,}}
\newcommand{\dsl}{\pa \kern-0.5em /}

\font\mybb=msbm10 at 12pt
\def\bb#1{\hbox{\mybb#1}}

\def\bC {\bb{C}}

\def\appendix#1{
  \addtocounter{section}{1}
  \setcounter{equation}{0}
  \renewcommand{\thesection}{\Alph{section}}
  \section*{Appendix \thesection\protect\indent \parbox[t]{11.15cm}
  {#1} }
  \addcontentsline{toc}{section}{Appendix \thesection\ \ \ #1}
  }

\begin{document}

\begin{flushright}
\small
DAMTP-2007-99\\
{\bf arXiv:0710.5709}\\
\date \\
\normalsize
\end{flushright}
\thispagestyle{empty}

\begin{center}


\vspace{.7cm}

{\Large {\bf  From Wave Geometry to Fake Supergravity$^*$\\}}

\bigskip

\vspace{1.2cm}

{\large Paul K. Townsend}

\vskip 1truecm

{Department of Applied Mathematics and
Theoretical Physics\\
Centre for Mathematical Sciences, University of Cambridge\\
Wilberforce Road, Cambridge, CB3 0WA, UK}
\vspace{.7cm}


\vskip2cm
\Large{{\bf Abstract}}

\end{center}

\begin{quotation}

The  `Wave Geometry'  equation of the pre-WWII Hiroshima program is also the key equation of  
the current `fake supergravity'  program. I review the status of  (fake) supersymmetric
domain walls and (fake) pseudo-supersymmetric cosmologies. An extension of the
domain-wall/cosmology correspondence to a triple correspondence  with instantons shows 
that `pseudo-supersymmetry'  has another  interpretation as Euclidean supersymmetry.

\end{quotation}

\vfill
\vfill
\vfill
\vfill
\vfill
\hrule width 5.cm
\vfill
{\small
\noindent $*$ Contribution to the proceedings of the 5th International 
Symposium on Quantum Theory and Symmetries (QTS5), Vallodolid, July 2007. 
\\ }

\newpage
\setcounter{page}{1}

\section{Introduction}
\setcounter{equation}{0}

The atomic bomb dropped on Hiroshima at the end of WWII killed many members of a group of physicists 
who were pursuing a `Wave Geometry'  (WG) program that was, in some respects, a precursor to supergravity. Although supersymmetry was then unknown,  the Hiroshima physicists were inspired by the Dirac  equation as the `square root' of the wave equation and aimed to do something similar for gravity. Their results were published in book form in 1962 \cite{Mimura:1962}. Central to the WG program was 
the equation
\be\label{WG}
\left(D_\mu + M_\mu \right) \kappa=0\, , \quad (\mu=0,1,2,3)
\ee
where $\kappa$ is a spinor wave-function, $D_\mu$ is the usual covariant derivative on spinors, and $M_\mu$ are matrix-valued functions. Much effort went into an attempt to determine, for all possible choices of the matrix functions $M_\mu$, the spacetime metrics that would allow a non-zero $\kappa$. In effect, the WG program involved a  classification of  four-dimensional spacetimes admitting spinors that are covariantly constant with respect to some  connexion with values in $Gl(4;\bC)$. This technical aspect of the program continues today in the effort to determine all supersymmetric solutions of supergravity theories, in particular the 10 and 11-dimensional supergravity theories of relevance to string/M-theory, and also in the ideas of  `fake supergravity' (FS) some of which will be reviewed here. 

The WG  equation  arises naturally in gauged supergravity theories from the requirement of partial preservation of supersymmetry since the vanishing of the local supersymmetry variation of the gravitino field for some non-zero spinor parameter $\kappa$ yields precisely an equation of the above form. In this context the WG equation is usually called a `Killing spinor'  (KS) equation. Its integrability conditions  are sometimes called `BPS' equations because they are first-order equations that are analogous to first-order equations introduced in the context of field theory solitons  by Prasad and Sommerfield \cite{Prasad:1975kr} and Bogomolnyi \cite{Bogomol'nyi:1976}.  Although there is often a natural relation of such equations to supersymmetry, any consequence of supersymmetry for the classical theory must be independent of the fermions and may therefore be applicable to a much wider class of theories. This 
was the general idea behind a derivation of first-order equations for $D$-dimensional domain wall spacetimes in \cite{Skenderis:1999mm}, which assumed a `supergravity-inspired' form of the scalar field potential introduced previously \cite{Townsend:1984iu}.  In this context, one considers equations of the WG form with $M_\mu\propto \Gamma_\mu$, where $\Gamma_\mu$ ($\mu=0,1,\dots,D-1$) are a set of Dirac matrices. The associated methods  currently go by the name of `fake supergravity' \cite{Freedman:2003ax}, which suggests compatibility with `genuine' supergravity although this aspect has been (more or less) fully understood only recently. Here I show how the supersymmetric domain walls of  $N=1$ $D=4$ supergravity \cite{Cvetic:1992bf,Cvetic:1996vr}  can be found from the FS formalism, following steps spelled out  in \cite{Behrndt:2000tr,Celi:2004st,Sonner:2007cp,Skenderis:2007sm}.

The fake supergravity formalism can be extended to cosmology via the `Domain-Wall/Cosmology correspondence': this is the observation that any homogeneous and isotropic cosmological solution of a model with potential $V$ can be obtained by analytic continuation of a domain wall solution of the same model but with potential $-V$  \cite{Cvetic:1994ya,Skenderis:2006jq}. A further feature of this correspondence is that a domain wall admitting Killing spinors corresponds to a cosmological solution admitting `pseudo-Killing' spinors \cite{Skenderis:2006jq}. Both Killing spinors and pseudo-Killing spinors are non-zero solutions of a WG equation of the form (\ref{WG}), but they differ according to the hermiticity properties of the `mass-matrix' $M=\Gamma^\mu M_\mu$.  If $M$ is hermitian (anti-hermitian) then we say that  a non-zero solution $\kappa$ is a Killing (pseudo-Killing) spinor. Here, by using results of
\cite{Bergshoeff:2005cp},  I  extend the Domain-Wall/Cosmology  correspondence to a triple correspondence between domain-walls, cosmologies and `cosmological instantons'.  This shows that the notion of a  pseudo-supersymmetric cosmology is closely related to the (possibly more familiar) notion of a supersymmetric instanton.

Also reviewed here, in the restricted context of flat domain walls or cosmologies, is the remarkable connection \cite{Skenderis:2006jq,Skenderis:2006rr} of the FS formalism to Hamilton-Jacobi (HJ) theory. As shown in \cite{Salopek:1990jq,de Boer:1999xf}, one can `reduce' the HJ equation for the `principal' function to a simpler equation for a function that depends only on the scalar fields; this is analogous to the reduction to an equation for Hamilton's characteristic function for a particle in a time-independent potential. This `reduced HJ equation is just the equation for the potential in terms of the superpotential, and the associated BPS equations are equivalent to the usual first-order equations of  the HJ formalism.

\section{Domain Walls and Fake Supersymmetry}
\setcounter{equation}{0}

We shall focus here on {\it flat} walls. By this we mean that the $(D-1)$-dimensional `worldvolume' geometry of the wall is Minkowski (since we consider only static walls).  To maintain the generic isometries of
such a metric, all fields other than scalar fields must vanish, 
so the general framework for a study of gravitational domain walls 
is gravity coupled to some number $n$ of scalar fields $\phi^i$ 
($i=1,\dots,n$) taking values  in a Riemannian `target' space with 
metric $G_{ij}$ and with potential  energy function $V(\phi)$. 
The Lagrangian density for such a model takes the form 
\be\label{lagmulti}
{\cal L} = \sqrt{-\det g} \left[{\cal R} - \frac{1}{2}G_{ij}\, 
\partial\phi^i\cdot \partial \phi^j  - V\right]\, , 
\ee
where $g$ is the metric for a $D$-dimensional spacetime and ${\cal R}$ is its scalar curvature. 

Introducing the $D$-dependent constants
\be
\alpha = \sqrt{(D-1)/[2(D-2)]}\, ,\qquad 
\beta = 1/\sqrt{2(D-1)(D-2)}\, , 
\ee
we may put the spacetime metric for a {\it flat} domain wall in the form 
\be
ds^2_D =   \left(e^{\alpha\varphi} f \right)^2 dz^2 + e^{2\beta\varphi}\left[-d\tau^2 +  d{\bf x}\cdot d{\bf x}\right]\, , 
\ee
where ${\bf x}$ are cartesian coordinates on the wall, and $\varphi$ and $f$ are functions of $z$. The inclusion of the function $f$ ensures that we maintain 
$z$-reparametrization invariance, so a choice of $f$ amounts to a choice of parametrization. 
The `standard' choice is
\be\label{stand}
f= e^{-\alpha\varphi}\, , 
\ee
because the parameter $z$ is then an affine distance parameter.

Taking the scalar fields to be functions of $z$ only, we have the  following effective Lagrangian for the variables $(\varphi, \{\phi\})$: 
\be\label{Leff1}
L_{eff} = \frac{1}{2} f^{-1} \left(\dot\varphi^2 - |\dot\phi|^2 \right) - f e^{2\alpha\varphi} V\, ,  
\ee
where $|..|$ is the norm in the target space metric.  If we now assume that there exists a real function $V$ such that \cite{Townsend:1984iu}
\be\label{fakeV}
V= 2\left[ G^{ij}\partial_i W\partial_j W - \alpha^2 W^2\right] \, , 
\ee
then we may rewrite $L_{eff}$ as \cite{Skenderis:1999mm}
\be
L_{eff} = \frac{1}{2}f^{-1}\left( \upsilon_\pm^2 - |{\rm v}_\pm|^2\right) \pm \frac{d}{dz} 
\left( 2e^{\alpha\varphi} W\right)\, , 
\ee
where
\be
\upsilon_\mp= \dot\varphi  \mp 2\alpha fe^{\alpha\varphi}W\, , \qquad
{\rm v}^i_\pm = \dot\phi^i \pm 2fe^{\alpha\varphi}G^{ij}\partial_j W\, . 
\ee
This form of $L_{eff}$ shows that  the equations of motion will be satisfied by any solution of the first-order equations 
\be\label{upv}
\upsilon_\mp =0\, , \qquad {\rm v}^i_\pm = 0 \qquad (i=1,\dots,n),
\ee
for {\it either} the upper sign {\it or} the lower sign.   Moreover, the Hamiltonian constraint obtained from the variation of $f$ can be put into the form
\be
\upsilon_-{\rm v}^i_+ = \upsilon_+ {\rm v}^i_-\, , 
\ee
and this is clearly satisfied too. 

For the standard gauge choice (\ref{stand}) the equations (\ref{upv})  become
\be\label{BPSfixed}
\dot\varphi = \pm 2\alpha W\, , \qquad \dot\phi^i = \mp 2G^{ij}\partial_j W\, . 
\ee
This derivation of these equations is similar to Bogomolnyi's derivation of first-order equations for field theory solitons in models that  are supersymmetrizable;  in the context of the supersymmetric theory, these `Bogomolnyi' equations are just the BPS equations for  supersymmetry preservation.  Is there a similar interpretation for  the equations (\ref{BPSfixed})?  Consider the `supergravity-inspired' Killing spinor equation \cite{Skenderis:1999mm,
Freedman:2003ax,Sonner:2005sj}
\be\label{KS}
\left(D_\mu -\alpha\beta W\Gamma_\mu\right)\kappa =0 \, ,
\ee
where $D_\mu$  is the standard covariant derivative on spinors and 
$\Gamma_\mu$ are Dirac matrices ($\mu=0,1,\dots,D-1$).  
It can be shown that the integrability conditions are 
\be\label{BPSkil}
\dot\varphi = \pm 2\alpha W \, , \qquad |\dot\phi|^2 = \mp
2\dot\phi^i\partial_i W\, ,  
\ee
and the Killing spinors are 
\be
\kappa(z) = e^{\frac{1}{2}\beta\varphi(z)}\kappa_0 \, 
\qquad \Gamma_z{\kappa}_0 = \pm {\kappa}_0\, . 
\ee

For a single scalar field, the equations (\ref{BPSkil}) are equivalent to (\ref{BPSfixed}) whenever $|\dot\phi|$ is non-zero, but for multi-scalar models we need additional equations. 
Consider the following `supergravity-inspired' conditions. 
\be\label{extraeq}
\left(\Gamma^\mu \partial_\mu \phi^i + 2G^{ij}\partial_j W 
\right)\kappa =0\, \qquad (i=1,\dots,n). 
\ee
Using the results that follow from the existence of Killing spinors, one finds that there are no further conditions on $\kappa$ provided that
\be
\dot\phi^i = \mp 2G^{ij}\partial_j W\, . 
\ee
Thus, the equations (\ref{KS}) and (\ref{extraeq}) together imply the first-order equations 
(\ref{BPSfixed}). This fact prompted the authors of  \cite{Freedman:2003ax} to refer to any model of gravity coupled to scalars for which the scalar potential takes the form (\ref{fakeV}) (or the generalization needed for curved domain walls) as a `fake supergravity' theory.  Note that there is no obvious relation of  `fake'  to `genuine' supergravity (except for $D=3$) so the conditions  (\ref{KS}) and (\ref{extraeq}) for `fake supersymmetry' are not related in any obvious way to conditions for preservation of `genuine' supersymmetry.  Nevertheless, one would hope that  the requirement of fake  supersymmetry is consistent with the requirement of supersymmetry when  the FS  theory happens to be the bosonic truncation of  a `genuine' supergravity theory. Let us now consider this issue  before moving on to cosmology.  

\subsection{Consistency with supergravity}

One source of the `inspiration'  for  FS  is minimal $D=3$ supergravity coupled to scalar supermultiplets because the potential $V$ in this case is given precisely by (\ref{fakeV}), and the same is true of the Killing spinor equation. The relation of fake to `genuine' supergravity for $D>3$ (when such a theory exists) is not so obvious. Domain wall solutions of minimal $D=4$ supergravity coupled to chiral supermultiplets have been much studied  \cite{Cvetic:1992bf,Cvetic:1996vr}. Let  us now see how the FS formalism  emerges in this context. The scalar fields are the complex first components of the chiral superfields, so the number $n$ or real scalar fields is even. Let $\Phi^\alpha$ ($\alpha=1,\dots n/2$) be these complex scalar fields, which must parametrize a  K\"ahler `target space'. The target space metric takes the form
$G_{\alpha\bar\beta}= \partial_{\alpha}\partial_{\bar\beta}{\cal K}$, 
where ${\cal K}(\Phi,\bar\Phi)$ is the  (real) K\"ahler potential. 
The scalar potential for this model is
\be
V= \frac{1}{2}e^{\cal K} \left[ G^{\alpha\bar\beta} 
D_{\alpha}P D_{\bar\beta}\bar P - 3 |P|^2\right]
\ee
where $P(\Phi)$ is the holomorphic superpotential, and  
\be
D_\alpha P = \partial_\alpha P + \left(\partial_\alpha {\cal K}\right) P
\ee  
is  the K\"ahler gauge-covariant derivative. When 
$V$ is expressed in terms of the real  `superpotential'
\be\label{realSP}
W= e^{{\cal K}/2} |P|\, ,  
\ee
it takes the form \cite{Behrndt:2000tr,Skenderis:2007sm}
\be\label{multi}
V= 2 \left[G^{\alpha\bar\beta} \partial_\alpha W\partial_{\bar\beta}W
  -  (3/4) W^2\right]\, , 
\ee
which agrees with (\ref{fakeV}). 

Now let $\Gamma_\mu$ ($\mu=0,1,2,3$) be real Dirac matrices 
and define $\gamma_5 = \Gamma_0\Gamma_1\Gamma_2\Gamma_3$, 
which is also real and squares to minus the identity matrix.  In terms of 
a pair of complex conjugate (anti)chiral spinors $\kappa_\pm$ (eigenspinors of $i\gamma_5$
with eigenvalues $\pm1$) the supergravity Killing  spinor equation is
\be
\left(D_\mu + \frac{i}{2}A_\mu\right)\kappa_+  - 
\frac{1}{4}e^{{\cal K}/2} P \Gamma_\mu \kappa_- =0
\ee
where \be\label{Kcon}
A= -\frac{i}{2}\left(d\Phi^\alpha \partial_\alpha {\cal K} - 
d\Phi^{\bar\beta}\partial_{\bar\beta}{\cal K}\right)
\ee
is the K\"ahler connection one-form.   In terms of the new real spinor parameter 
\cite{Skenderis:2007sm}
\be
\kappa =\left(\bar P/P\right)^{1/4} \kappa_+ 
+ \left(P/\bar P\right)^{1/4} \kappa_-\, , 
\ee
one finds that the Killing spinor equation simplifies to
\be
\left(D_\mu + \frac{i}{2}\tilde A_\mu - \frac{1}{4} W \Gamma_\mu\right) \kappa=0\, ,
\ee
where $\tilde A$ is given by (\ref{Kcon}) but with ${\cal K}$ replaced 
by $\tilde{\cal K} = 2\log W$.  For co-dimension one metrics the
connection $\tilde A$ is necessarily pure gauge, so the integrability 
conditions for a $D=4$ Killing spinor in a domain wall background are 
the same as those for the fake Killing spinors defined by
(\ref{KS}). 

Finally, the remaining supersymmetry preservation
conditions of $D=4$ supergravity are 
\be
\Gamma^\mu\partial_\mu \Phi^\alpha \kappa_- + 2e^{{\cal K}/2}
G^{\alpha\bar\beta}D_{\bar\beta}\bar P \kappa_+ =0\, . 
\ee 
When these equations are expressed in terms of $W$ and the new spinor
parameter $\kappa$, one finds that
\be
\left(1+ i\gamma_5\right) \left[\Gamma^\mu \partial_\mu \Phi^\alpha + 4G^{\alpha\bar\beta}\partial_{\bar\beta}W \right]\kappa =0\, . 
\ee
On setting $\Phi^\alpha = A^\alpha + iB^\alpha$ for real scalar fields $(A^\alpha,B^\alpha)$,
one finds the equivalent condition
\be\label{sugrastuff}
\left[\Gamma^\mu\partial_\mu A^\alpha + 2G^{\alpha\bar\beta} \frac{\partial W}{\partial A^\beta}\right] \kappa
= \gamma_5 \left[\Gamma^\mu\partial_\mu B^\alpha + 2G^{\alpha\bar\beta} \frac{\partial W}{\partial B^\beta}\right] \kappa\, . 
\ee
This is not yet (\ref{extraeq}) but we should now recall that for domain walls the only non-zero derivatives of the scalar fields are the $\partial_z$ derivatives, and also that  $\kappa$ is an eigenspinor of 
$\Gamma_z$ for a supersymmetric domain wall. Since $\Gamma_z$ anti-commutes with $\gamma_5$, multiplying both sides of  (\ref{sugrastuff}) by $\Gamma_z$ effectively yields the same equation 
but with opposite sign for the right  hand side, so both left and right hand sides must vanish separately 
and the consequences of (\ref{sugrastuff}) are therefore the same as those of (\ref{extraeq}).

\section{Cosmology and pseudo-supersymmetry}
\setcounter{equation}{0}

The relevant Lagrangian density  for homogeneous and isotropic cosmologies is again (\ref{lagmulti}) because the symmetries again permit only scalar fields to be non-zero. The spacetime $D$-metric for a {\it flat} homogeneous and isotropic universes may be put into the form
\be
ds^2_D =   -\left(e^{\alpha\varphi} f \right)^2 dt^2 + a^2(t) \, d{\bf X}\cdot d{\bf X}
\ee
where ${\bf X}$ are cartesian coordinates for a $(D-1)$-dimensional Euclidean space, and 
\be
a= e^{\beta\varphi (t)}
\ee
is the cosmological scale factor. A choice of $f$ amounts to a choice of time parametrization. 
The `standard' choice is
\be\label{standcos}
f= e^{-\alpha\varphi}\, , 
\ee
because $t$ is then the standard Friedmann-Lemaitre-Robinson-Walker (FLRW) time. For this choice, the inverse Hubble length is
\be\label{Hubb}
H= \dot a/a \equiv  \beta \dot \varphi\, ,  
\ee
where the overdot now indicates differentiation with respect to $t$. 

One might suppose  that cosmologies, in contrast to domain walls,  cannot be supersymmetric because they are time-dependent. While this is true in
the context of standard supergravity theories, there are (non-unitary)
`pseudo-supergravity' theories for which some cosmological solutions
{\it are} supersymmetric \cite{Bergshoeff:2007cg,Skenderis:2007sm,Vaula:2007jk}. 
These theories can be viewed as different real forms of a complexified theory with holomorphic action functional  \cite{Bergshoeff:2007cg}, and complex superpotential $W$. 
Variant real forms with Lorentzian signature spacetime are possible only 
for extended supergravities because the analytic continuation also modifies the reality 
properties of spinors.  For example, for $D=4$ one has $N$ chiral gravitino fields that are related 
to their complex conjugates by some  linear transformation that is trivial for standard $N$-extended supergravity but `twisted'  for variant supergravities, in a way that requires $N$ to be even
\cite{Pilch:1985aw,deWit:1987sn}. 

However, we may extend the notion of a 
`supersymmetric'  cosmology of a pseudo-supergravity theory to a fake pseudo-supersymmetric cosmology of a fake  pseudo-supergravity theory, and in this context we may allow all 
`pseudo-Killing'  spinors to be complex.   In fact, this is the context  in which  the idea of 
a pseudo-Killing spinor first arose \cite{Skenderis:2006jq}. We may also allow any spacetime 
dimension $D$,  and any superpotential  function $W$, but the analytic continuation takes $W\to iW$,
where the `new' $W$ is again real. As a result, the scalar potential is again given by (\ref{fakeV}) but {\it with the opposite sign}: 
\begin{equation}\label{pseudoV}
V= -2\left[ G^{ij}\partial_i W\partial_j W - \alpha^2 W^2\right] \, .
\end{equation}
Similarly, the Killing spinor equation (\ref{KS}) becomes the pseudo-Killing spinor equation
\begin{equation}\label{pseudoKS}
\left(D_\mu - i\alpha\beta W\Gamma_\mu\right)\kappa =0 \, . 
\end{equation} 
The pseudo-Killing spinors take  the form
\begin{equation}
\kappa = \sqrt{a(t)} \, \kappa_0\, ,  \qquad i\Gamma_t \kappa_0= \pm \kappa_0\, , 
\end{equation}
which shows that the pseudo-Killing spinor is a `square-root' of the cosmological scale factor. 
Note that $\Gamma_t$ squares to minus the identity, so the factor of $i$ in the projection is needed
for consistency; this factor is provided by the factor of $i$ in (\ref{pseudoKS}).

Let us now restrict attention to models with a single scalar field $\sigma$, which may be viewed as the 
inflaton field of inflationary cosmology. Assuming that $\sigma$ is a monotonic function of time, one finds that  the integrability conditions for pseudo-Killing spinors are
\be\label{intsingle}
\dot\varphi = \pm 2\alpha W\, , \qquad \dot\sigma = \mp 2W'\, , 
\ee
where $W'=dW/d\sigma$. The first of these equations is just (\ref{Hubb}) with
\be
H= \pm 2\alpha\beta W\, , 
\ee
so the inverse Hubble length viewed as a function of $\sigma$ is proportional to the superpotential. 
Remarkably, it appears that it is $W(\sigma)$ rather than $V(\sigma)$ that is most directly constrained 
(for $D=4$, naturally)  by  the data \cite{LAV:2007}.  

Except for the different interpretation of the independent variable, the equations (\ref{intsingle}) are just the specialization to a single scalar of the first-order equations (\ref{BPSfixed}) for domain walls. This is one aspect of the general Domain-Wall/Cosmology correspondence \cite{Cvetic:1994ya,Skenderis:2006jq}.  
As will now be explained, this is   actually part of a larger correspondence that involves cosmological instantons.

\subsection{The Domain-wall/Cosmology/Instanton correspondence}

For simplicity we continue to consider the single-scalar model, but 
to allow consideration of Euclidean-signature metrics we also 
introduce a sign $\epsilon$ so that  the Lagrangian density is 
\cite{Bergshoeff:2005cp}.
\be
{\cal L} = \sqrt{\epsilon \det g} \left[R - 
\frac{1}{2}\left( \partial\sigma\right)^2- V\right]\, , 
\ee
where $\epsilon=1$ for Euclidean signature and $\epsilon=-1$ for  
Lorentzian signature.   Our interest here is with co-dimension one configurations, and for Lorentzian signature this means that we are considering either domain-wall  or  (homogeneous and isotropic) cosmological spacetimes; we pass over the possibility of a foliation by null hypersurfaces.   Let us  introduce a new sign $\eta$ such that $d\Omega^2_\eta$ is the $SO(D-1)$-invariant metric on the unit radius  $(D-2)$-sphere if $\eta=-1$, and the $SO(1,D-3)$-invariant metric on the unit  radius $(D-2)$-hyperboloid if $\eta=1$. Now consider a metric of the form
\be
ds^2_D = -\epsilon\eta\,  \left(e^{\alpha\varphi} f \right)^2 dz^2 + e^{2\beta\varphi}\left[-\eta  \frac{dr^2}{1+\eta k r^2} + r^2d\Omega_\eta^2 \right]\, , 
\ee
where $\varphi$ and $f$ are functions of $\tau$, and $k$ is either zero or $\pm1$. The $D$-dependent constants $\alpha$ and $\beta$ are as given in (\ref{fakeV}) and (\ref{KS}).  For $\epsilon=-1$, the choice $\eta=-1$ yields the metric of a homogeneous and isotropic cosmology, describing a universe that is closed if $k=1$, open if $k=-1$ and flat if $k=0$. 

Again for $\epsilon=-1$ but now setting $\eta=1$ we have the metric of a domain wall;  its worldvolume geometry is anti-de Sitter if $k=-1$, de Sitter if $k=1$ and Minkowski if $k=0$.  In this case $z$ is a space coordinate parametrizing distance from some fiducial leaf of the foliation that can be viewed, for fixed $r$ (which is now a time coordinate) as `the domain wall'.  Finally, when $\eta=1$ we may choose $\epsilon=1$ to get a Riemannian metric, which may be (loosely) interpreted as an `instanton'. In all cases we must choose the scalar field $\sigma$ to be a function only of $z$ in order to preserve the (generic) isometries of the metric.
 
Because we have maintained $z$-parametrization invariance by the inclusion of the function $f$ in the ansatz, it is legitimate to substitute the ansatz into the action. After integration by parts and a rescaling, this yields  the effective Lagrangian 
\be
L_{eff} = \frac{1}{2} f^{-1} \left(\dot\sigma^2 - \dot\varphi^2 \right) -  (\epsilon\eta) f e^{2\alpha \varphi}\, V_{eff}
\ee
where the overdot indicates differentiation with respect to $z$, and
\be
V_{eff} (\sigma,\varphi) = V(\sigma)  - \frac{k}{2\beta^2} e^{-2\beta\varphi}\, . 
\ee
Note that  the effective Lagrangian depends on the signs $\epsilon$ and $\eta$ only through their product, and that it is invariant under $\epsilon\eta \to -\epsilon\eta$ provided that $V\to -V$ and $k\to -k$.  Setting $\epsilon=-1$, we deduce the domain-wall/cosmology correspondence: {\it for every domain wall solution of a model with potential $V$ there is a cosmology of the model with potential $-V$ (with opposite sign $k$ if $k\ne0$), and  vice-versa}.

Recall that the choice $\epsilon=1$ and $\eta=-1$, and hence $\epsilon\eta=-1$, yields solutions with a Riemannian metric, but $\epsilon\eta=-1$ for domain-walls too,  so each solution of the effective equations of motion for a model with potential $V$ yields both a domain wall solution of the Lorentzian-signature Einstein-scalar equations {\it and} a solution of the Euclidean-signature Einstein-scalar equations. The latter can be interpreted as an instanton, but of the model with potential 
$-V$ because instanton solutions of a mechanical model are precisely solutions with a flipped sign of the potential. We thus have the following extension of the domain-wall/cosmology correspondence: {\it to each domain wall solution of a model with potential $V$ there corresponds both a cosmology and an instanton of the model with potential  $-V$ (although the latter is actually found from the effective Lagrangian with potential $V$)}. 

Note that this correspondence holds quite generally, irrespective of whether the solutions are (pseudo)supersymmetric. However, there will be a correspondence between supersymmetric domain walls, pseudo-supersymmetric cosmologies and  `supersymmetric'  cosmological instantons, which will admit Killing spinor solutions of (\ref{KS})  but  for a Euclidean spacetime metric. This metric could also be viewed as a domain wall with a Euclidean worldvolume metric; in any case, the integrability conditions are the 
same as those for a domain wall.

\section{Hamilton-Jacobi and supersymmetry}
\setcounter{equation}{0}

Returning to the case of cosmologies ($\eta=-1$) and domain walls ($\eta=1$) with an arbitrary number of scalar scalar fields,  we introduce new variables $(\pi,p_i)$ canonically conjugate to the variables $(\varphi,\phi^i)$, and pass to the Hamiltonian form of the effective Lagrangian, which is
\begin{equation}
L_0 = \dot\varphi \pi + \dot\phi^i p_i - f {\cal H}
\end{equation}
where 
\begin{equation}
{\cal H} = \frac{1}{2}\left(\pi^2 -G_{ij}p^ip^j\right) + \eta\, e^{2\alpha\varphi} V_{eff}\, . 
\end{equation}
Note that $f$ is now a Lagrange multiplier for the Hamiltonian
constraint ${\cal H}=0$. The Hamilton-Jacobi
(HJ) equation for this system is found by setting
\begin{equation}
\pi = \frac{\partial S}{\partial\varphi}\, , \qquad p_i= \frac{\partial S}{\partial \phi^i}\, , 
\end{equation}
for Hamilton's principal function $S(\varphi,\{\phi\})$, in which case 
the Hamiltonian constraint becomes
\begin{equation}
\left(\frac{\partial S}{\partial \varphi}\right)^2 -
G^{ij}\left(\frac{\partial S}{\partial \phi^i}\right)
\left(\frac{\partial S}{\partial \phi^j}\right) +
2\eta\,  e^{2\alpha\varphi} \, V_{eff} =0\, . 
\end{equation}
This is the HJ equation. From a solution of this equation one finds a 
solution of the equations of motion via the first-order equations
\begin{equation}\label{HJfirst}
f^{-1}\dot \varphi = \frac{\partial S}{\partial \varphi}\, , \qquad f^{-1}\dot\phi^i = - G^{ij}\frac{\partial S}{\partial\phi^j}\, . 
\end{equation}

We shall focus here on the $k=0$ case for which $V_{eff}=V$. In this case, 
the HJ equation is solved by 
\begin{equation}\label{HJsoln}
S= \pm 2 e^{\alpha\varphi} W(\{\phi\})\, . 
\end{equation}
provided that the function $W$ satisfies
\begin{equation}\label{reducedHJ}
 V= 2\eta\left[G^{ij}\frac{\partial W}{\partial \phi^i} 
\frac{\partial W}{\partial \phi^j} 
 -\alpha^2 W^2\right]\, . 
\end{equation}
Given a solution of this equation for $W$ we now get a solution of 
the equations of motion via the `reduced' first-order equations 
\begin{equation}\label{redfirst}
f^{-1}\dot\varphi = \pm 2\alpha e^{\alpha\varphi} \, W\, , \qquad
f^{-1}\dot \phi^i = \mp 2 e^{\alpha\varphi} \, 
G^{ij}\frac{\partial W}{\partial \phi^j} \, . 
\end{equation}
In the `gauge' $f= e^{-\alpha\varphi}$, for which the independent 
variable is an affine parameter, and for a single scalar $\sigma$, 
these equations reduce to (\ref{intsingle}), which can be used 
to construct the function $W$ from any given solution for which the 
function $\dot\sigma(z)$ has no zeros \cite{Freedman:2003ax,Sonner:2005sj};
the construction yields a multi-valued  $W$ in those cases for which $\dot\sigma(z)$ 
has isolated zeros. The remaining  cases, such as domain walls asymptotic to unstable adS 
spactimes \cite{Skenderis:2006rr},  necessarily correspond to some 
other type of solution of the HJ equation and will not be discussed here. 

Recall that $\eta =-1$ for cosmologies, in which case we have
recovered the `reduced HJ' equation of Salopek and Bond
\cite{Salopek:1990jq}. For domain walls we have $\eta=1$, and
we then recover the analogous reduced HJ equation of de Boer et
al. \cite{de Boer:1999xf}. Remarkably, these expressions for the
scalar potential in terms of the `reduced HJ' function $W$ are
precisely the same as the expressions given earlier for a fake
(pseudo)supergravity theory, which coincide with the
(pseudo)supergravity expressions whenever these are applicable. 
Moreover, the first-order equations (\ref{redfirst}) are precisely the 
`BPS' conditions of fake (pseudo)supersymmetry, which coincide with the
conditions required for preservation of 1/2 supersymmetry in
(pseudo)supergravity whenever this is applicable.

\section{Discussion}
\setcounter{equation}{0}

The notion of a spinor that is covariantly constant with respect to a connexion more general than 
the usual spin-connexion appears to have first arisen,  in physics if not also in mathematics, in the 
context of the `wave geometry' program of the pre-WWII Hiroshima group. As is well-known, the  study of such spinors arises naturally in the context of supergravity theories and is of relevance to studies of string/M-theory, but it is of more general relevance and  it is this fact that is exploited by  fake supergravity, which can perhaps be viewed as the intellectual descendent of  Wave Geometry. 

As we have seen, fake supergravity is consistent with standard $D=4$ $N=1$ supergravity, when applicable, but generalizes the `BPS' conditions associated with preservation of 1/2 supersymmetry by a domain-wall spacetime.  As we have also seen (at least for flat walls) there is a remarkable concordance between fake supergravity and Hamilton Jacobi theory: the BPS equations for a (fake) supersymmetric domain-wall solution are precisely the first-order equations of HJ theory while the relation between the potential and the superpotential is the HJ equation.
It is amusing to note that the history of HJ theory was reversed for FS theory, in the following sense: when Hamilton found the equation for his `principal'  function that we now call the Hamilton-Jacobi equation, he deduced it from the equations of motion, i.e. from Hamilton's equations. When Jacobi saw  Hamilton's paper, he realized the central importance of this equation,  and  that one could reverse the logic to find a solution of Hamilton's equations from a solution of the HJ equation. In the fake supergravity case, Jacobi's perspective came first: the `reduced HJ eqution' was proposed as an equation to be solved for the superpotential given the potential, and one then found a `fake' supersymmetric solution via the BPS equations. Subsequently, it was realized that the converse is also true: given a solution of the equations of motion one can  construct a superpotential satisfying the `reduced HJ equation, such that the given solution is fake supersymmetric (i.e. solves the BPS equations). 

These results apply as much to cosmology as to to domain walls as a consequence of the Domain-Wall/Cosmology (DW/C) correspondence, although the notion of `supersymmetry' must be replaced by `pseudo-supersymmetry'. A new observation of this paper is that if all scalar fields are true scalars rather than pseudo-scalars,  then the DW/C correspondence extends to a triple correspondence with real solutions of the Euclidean theory with Euclidean metric, i.e. instantons\footnote{Complex instanton solutions would probably be needed to accomodate pseudo-scalars.}. Starting with the cosmological solution, one gets the instanton solution by a simple analytic continuation of the time coordinate. As shown in \cite{Bergshoeff:2005cp,Bergshoeff:2005bt} cosmologies and instantons may actually be part of the `same'  solution; such that, for example, a collapsing big-crunch universe  becomes an expanding big-bang universe by passing through a phase in which the metric has Euclidean signature  \cite{Russo:2004am}, so it is perhaps not so surprising that both have a corresponding domain-wall solution.

The concordance between the FS and HJ formalism might seem surprising when one reflects on the difference between a static supersymmetric  domain wall and the apparent time-dependence implicit in the effective mechanical model.  Of course, one should interpret  `time' as space,  but precisely for this reason  the potential has the `wrong' sign relative  to the original gravity-scalar model, and this again makes  the `supersymmetry'  puzzling. One might suppose that the resolution of this puzzle is simply that there is no reason to think of the mechanical model as `supersymmetric'  because the `supersymmetry'  in question is just the statement that the {\it spacetime} admits Killing spinors, and this has no obvious implications at the level of the effective Lagrangian. However, the cosmological perspective suggests a different resolution. 

The same effective action that governs domain-wall solutions of the model with potential $V$ governs 
cosmological solutions of the model with potential $-V$, so in this case there is no sign flip of the potential in passing to the effective Lagrangian (as expected because the `time' of the effective mechanics is now cosmological time). While it might seem unlikely that a time-dependent solution
of this mechanical model could be `supersymmetric' in any reasonable sense, this is now no more unlikely than that the cosmological solution should be supersymmetric in the context of  some supergravity theory, and we know that some cosmologies {\it are} supersymmetric solutions of 
pseudo-supergravity theories. This suggests that there might exist a {\it pseudo-supersymmetric} extension of the mechanical model such that the BPS equations of this model are precisely the first-order equations of the HJ formalism as applied to cosmology.  As shown elsewhere  \cite{pkt:2007}, such a `pseudo-supersymmetrization'  is not only possible but can be used to re-derive the HJ  formalism!

\subsection*{Acknowledgments}
I am grateful to my collaborators Kostas Skenderis, Julian Sonner and Antoine Van Proeyen for all the discussions during the joint work on which this contribution is based. I am also grateful to  Gary Gibbons for pointing out to me the relevance of the Wave Geometry program to fake supergravity. 
This work was supported by an EPSRC Senior Research Fellowship.

\end{document}